\documentclass{PoS}

\usepackage{graphicx}
\usepackage{sidecap}
\usepackage{wrapfig}
\usepackage{latexsym}
\usepackage{epsfig}
\usepackage[mathscr]{eucal}
\usepackage{amsfonts}
\usepackage{amscd}
\usepackage{cite}
\usepackage{array}
\usepackage{amssymb}
\usepackage{colordvi}
\usepackage[centertags]{amsmath}
\usepackage{enumerate}
\usepackage{graphicx}
\usepackage{booktabs}
\usepackage{theorem}
\usepackage[footnotesize]{caption}
\usepackage{soul}
\usepackage{url}
\usepackage{mcite}
\usepackage{slashed}
\usepackage{color}
\usepackage{ulem}
\usepackage{bbm}
\usepackage[utf8]{inputenc}
\newcommand{\newc}{\newcommand}
\newc{\be}{\begin{equation}}
\newc{\ee}{\end{equation}}
\newc{\bea}{\begin{eqnarray}}
\newc{\eea}{\end{eqnarray}}
\newc{\ol}{\overline}
\newc{\wt}{\widetilde}
\newc{\bs}{\boldsymbol}
\newc{\m}{\mathcal}
\newc{\la}{\langle}
\newc{\ra}{\rangle}

\newcommand{\beq}{\begin{eqnarray}}
\newcommand{\eeq}{\end{eqnarray}}
\newcommand{\bpmatrix}{\begin{pmatrix}}
\newcommand{\epmatrix}{\end{pmatrix}}
\newcommand{\ba}{\begin{array}}
\newcommand{\ea}{\end{array}}

\renewcommand{\ol}{\text{1l}}

\renewcommand{\eqref}[1]{Eq.~(\ref{#1})}

\newcommand{\bc}{\begin{center}}
\newcommand{\ec}{\end{center}}

\newcommand{\hour}{{~\text{h}}}

\title{Extended Higgs sector beyond the MSSM and the LHC}

\ShortTitle{Extended Higgs Sectors at the LHC}

\author{\speaker{Rui Santos}\thanks{I would like to thank the organisers for financial support.}\\
        Centro de F\'{\i}sica Te\'{o}rica e Computacional,
    Faculdade de Ci\^{e}ncias,\\
    Universidade de Lisboa, Campo Grande, Edif\'{\i}cio C8
  1749-016 Lisboa, Portugal, and  \\
  ISEL - Instituto Superior de Engenharia de Lisboa,\\
   Instituto Polit\'ecnico de Lisboa
 1959-007 Lisboa, Portugal.\\
        E-mail: \email{rasantos@fc.ul.pt}\color{black}}

\abstract{One Higgs was found. Are there more? In this work we discuss simple extension of the scalar sector
of the Standard Model (SM) used as benchmark models by ATLAS and CMS in the searches 
for new scalars at the LHC. We discuss how much the discovered 125 GeV Higgs at the LHC resembles the SM Higgs and how will our understanding of the Higgs nature improve at future electron-positron colliders.
Models with extended Higgs sectors provide very interesting scenarios, from the existence of charged Higgs bosons to 
CP-violating scalars that can be probed by the experimental collaborations at the LHC. Comparison between the rates in the different
models show that in some cases the models could be distinguished.}

\FullConference{An Alpine LHC Physics Summit (ALPS2018)\\
		15-20 April, 2018\\
		Obergurgl, Austria}

\begin{document}

\section{Introduction}

\noindent
After the discovery of the Higgs boson, the search for new scalars by the experimental groups at CERN, further motivated the study of extensions
of the Standard Model (SM). Besides supersymmetric models, the simplest extensions of the scalar sector of the SM
provide an excellent framework for the interpretation of many searches and to motivate new searches. 
In this work we discuss a few extensions of the scalar sector of the SM.  We will discuss how efficiently can the parameter space of 
these simple extensions be constrained through the measurements of the Higgs couplings
and how SM-like is the SM-like Higgs boson. We furthermore try to understand if these models can be distinguished if a new scalar is found. All
models have a limit where the 125 GeV Higgs looks exactly like the SM at tree level and if all other particles are very heavy
it will be hard to probe their existence. At this stage it is not clear if electroweak radiative corrections play an important role
in the model's phenomenology. In fact, although the corrections may change significantly the tree-level couplings of the
125 GeV Higgs, there are large regions of the parameter space of the models where they are small enough to be 
inside the predicted error for the future LHC Higgs couplings measurements. 

\noindent
If the future measurements of the 125 GeV Higgs couplings are compatible with the SM predictions with ever increasing precision,
the models will approach more and more their SM-like limit, where they are all very similar as we will show. Only if a new scalar 
is found can we start probing the different possibilities for the new models. If this is the case, some models show very interesting properties,
some of which are very characteristic of specific models. We will discuss some particularly interesting scenarios of selected models.

\section{Building the models}

\noindent
When building extensions of the SM there are some very general features which make the models comply with experimental results 
in a simpler way. However, if it is true that all models need to provide a 125 GeV Higgs that is not a pure CP-odd scalar, any other constraints
should only be seen as a guide to build the simpler extensions compatible with the experimental results. Among the most relevant are:
\begin{itemize}
\item

\noindent
The $\rho$ parameter, which is measured with great precision, can be written as a function of the $SU(2)_L$ Isospin $T_i$, the Hypercharge $Y$,
and the vacuum expectation value of the fields $v_i$, as
\begin{equation}
\rho =\frac{m_W^2}{m_W^2 \, \cos^2 \theta_W}=\frac{\sum_i [4T_i(T_i+1)-Y_i^2]\, |v_i|^2 \, c_i}{\sum_i 2 \, Y_i^2 |v_i|^2}
\end{equation}
where $c_i =1 (1/2)$ for complex (real) representations. The simplest representation with $\rho=1$ is the singlet. The next one
is the doublet and after that comes the septet. Hence, extended models with an arbitrary number of singlets and doublets
have  $\rho=1$ at tree-level. Extensions with any other representations will need some kind of fine-tuning to
comply with  $\rho=1$ at tree-level.

\item

\noindent
Tree-level flavour changing neutral currents (FCNC) are experimentally very constrained. Models
with more than one doublet can give rise to tree-level FCNC. This problem is usually fixed with the
introduction of ad-hoc discrete symmetries imposed both on the scalar and on the fermion fields.

\end{itemize}

\subsection{SM + singlet (RxSM and CxSM)}

\noindent
We now present a few of the simplest models that obey all the conditions above. 
The simplest extension of the scalar potential of the SM is the addition of either a real (RxSM)
or a complex (CxSM) singlet. The complex field $\mathbb{S} =  S + iA$ has zero isospin and zero 
hypercharge and therefore 
only enters the model via mixing with the scalar field from the SM doublet. The CxSM version
of the potential is invariant under a global $U(1)$ symmetry, softly broken by linear
and quartic terms,
\beq
V = \frac{m^2}{2} H^\dagger H + \frac{\lambda}{4} (H^\dagger
H)^2+\frac{\delta_2}{2} H^\dagger H |\mathbb{S}|^2 +
\frac{b_2}{2}|\mathbb{S}|^2+ \frac{d_2}{4} |\mathbb{S}|^4 +
\left(\frac{b_1}{4} \mathbb{S}^2 + a_1 \mathbb{S} +c.c. \right)
\, ,  \label{eq:VCxSM}
\eeq
with the fields defined as
\begin{equation}
H=\left(\begin{array}{c} G^+ \\
    \dfrac{1}{\sqrt{2}}(v+h+iG^0)\end{array}\right) \quad \mbox{and} \quad
\mathbb{S}=\dfrac{1}{\sqrt{2}}\left[v_S+s+i(v_A+ a)\right] \;,
\label{eq:fieldsCxSM}
\end{equation}
where $v\approx 246$~GeV is the vacuum expectation value (VEV) of the $h$ field and $v_S$ and $v_A$
are the VEVs of the real and imaginary parts of the complex singlet field, respectively. 
Imposing  invariance under  $\mathbb{S} \to \mathbb{S}^*$ (or $A \to -A$),  implies that $a_1$ and $b_1$ are real.

\noindent
The vacuum structure determines the number of stable particles (see discussion in~\cite{Coimbra:2013qq}).
In the broken phase, where all 3 VEVs are non-zero and the three CP-even scalars mix, the mass eigenstates
$H_i$ are obtained via the rotation matrix $R$, which we parametrize as
\beq
R =\left( \begin{array}{ccc}
c_{1} c_{2} & s_{1} c_{2} & s_{2}\\
-(c_{1} s_{2} s_{3} + s_{1} c_{3})
& c_{1} c_{3} - s_{1} s_{2} s_{3}
& c_{2} s_{3} \\
- c_{1} s_{2} c_{3} + s_{1} s_{3} &
-(c_{1} s_{3} + s_{1} s_{2} c_{3})
& c_{2}  c_{3}
\end{array} \right) \; ,
\label{eq:rotsinglet}
\eeq
where we have defined  $s_{i} \equiv \sin \alpha_i$ and $c_{i} \equiv \cos \alpha_i$, with the angles varying in the range
$-\pi/2 \le \alpha_i <  \pi/2 $ and the masses of the neutral Higgs ordered as $m_{H_1} \leq m_{H_2} \leq m_{H_3}$.
A detailed account of the models can be found in \cite{Coimbra:2013qq}.

\subsection{SM + doublet (2HDM and C2HDM)\label{sec:2hdm}}

\noindent
The potential for the real (2HDM) and complex (C2HDM~\cite{Ginzburg:2002wt}) versions of the 2-Higgs-Doublet model, 
is chosen to be invariant under the $\mathbb{Z}_2$ transformations
$\Phi_1 \to \Phi_1$ and $\Phi_2 \to -\Phi_2$,
\beq
V &=& m_{11}^2 |\Phi_1|^2 + m_{22}^2 |\Phi_2|^2 - m_{12}^2 (\Phi_1^\dagger
\Phi_2 + h.c.) + \frac{\lambda_1}{2} (\Phi_1^\dagger \Phi_1)^2 +
\frac{\lambda_2}{2} (\Phi_2^\dagger \Phi_2)^2 \nonumber \\
&& + \lambda_3
(\Phi_1^\dagger \Phi_1) (\Phi_2^\dagger \Phi_2) + \lambda_4
(\Phi_1^\dagger \Phi_2) (\Phi_2^\dagger \Phi_1) + \frac{\lambda_5}{2}
[(\Phi_1^\dagger \Phi_2)^2 + h.c.] \; .
\eeq
The 2HDM is defined with all parameters and VEVs real, while the C2HDM
is built with real VEVs, but  $m_{12}^2$ and $\lambda_5$ complex.
The particle spectrum of the 2HDM include two CP-even scalars,
one CP-odd scalar and two charged Higgs. The C2HDM has
two charged scalars and three neutral scalar bosons with no definite CP
$H_i$   ($i=1,2,3$), ordered by ascending mass according to $m_{H_1} \le m_{H_2} \le m_{H_3}$. 
In the C2HDM the neutral mass eigenstates are obtained via the rotation of a matrix we again parametrise as 
$R$ in (~\eqref{eq:rotsinglet}\color{black} ), with the same allowed range for the mixing angles.
The 2HDM has 8 independent parameters while the C2HDM has 9 free parameters. For both models we define 
the common parameters $v = \sqrt{v_1^2 + v_2^2}  \approx 246\mbox{ GeV}$ and $\tan \beta=v_2/v_1$. 
The remaining free parameters for the 2HDM are $\alpha, \, m_{h}, \, m_{H}, \, m_{A}, \, m_{H^\pm} \,
\mbox{and} \, m_{12}^2 \;$, where $\alpha$ is the rotation angle in the CP-even sector.
For the C2HDM the remaining free parameters are $\alpha_{1,2,3}, \, m_{H_i}, \, m_{H_j}, \, m_{H^\pm} \, 
\mbox{and} \, Re(m_{12}^2)$. The third neutral Higgs mass is obtained from the other parameters \cite{ElKaffas:2007rq}. 
The 2HDM and C2HDM discussed in this work have no tree-level FCNCs due to 
the global $\mathbb{Z}_2$ symmetry imposed on the scalar doublets which is extended to the fermions
leading to the four independent Yukawa versions of the model: Type I, Type II, Flipped and Lepton Specific.  
All couplings for the C2HDM  can be found in~\cite{Fontes:2017zfn}.

\subsection{SM + doublet + singlet (N2HDM)\label{sec:n2hdm}}

\noindent
The potential chosen for the N2HDM~\cite{Muhlleitner:2016mzt} is  invariant under the $\mathbb{Z}_2$ symmetries
\begin{align}
  \Phi_1 \to \Phi_1\;, \quad \Phi_2 \to - \Phi_2\;, \quad \Phi_S \to \Phi_S \label{eq:2HDMZ2}
\end{align}
which is softly broken by the $m_{12}^2$ term and
\begin{align}
  \Phi_1 \to \Phi_1\;, \quad \Phi_2 \to \Phi_2\;, \quad \Phi_S \to -\Phi_S \label{eq:singZ2}
\end{align}
broken spontaneously by the singlet VEV.
We write the potential as
\beq
V &=& m_{11}^2 |\Phi_1|^2 + m_{22}^2 |\Phi_2|^2 - m_{12}^2 (\Phi_1^\dagger
\Phi_2 + h.c.) + \frac{\lambda_1}{2} (\Phi_1^\dagger \Phi_1)^2 +
\frac{\lambda_2}{2} (\Phi_2^\dagger \Phi_2)^2 \nonumber \\
&& + \lambda_3
(\Phi_1^\dagger \Phi_1) (\Phi_2^\dagger \Phi_2) + \lambda_4
(\Phi_1^\dagger \Phi_2) (\Phi_2^\dagger \Phi_1) + \frac{\lambda_5}{2}
[(\Phi_1^\dagger \Phi_2)^2 + h.c.] \nonumber \\
&& + \frac{1}{2} m_S^2 \Phi_S^2 + \frac{\lambda_6}{8} \Phi_S^4 +
\frac{\lambda_7}{2} (\Phi_1^\dagger \Phi_1) \Phi_S^2 +
\frac{\lambda_8}{2} (\Phi_2^\dagger \Phi_2) \Phi_S^2 \;.
\label{eq:n2hdmpot}
\eeq
This model is CP-conserving and has no dark matter candidate. The particle spectrum includes
two charged Higgs, one CP-odd boson and three CP-even scalars which again we denote
by $H_i$. One of the CP-even scalars is chosen to be the 125 GeV Higgs. The rotation from 
the gauge eigenstates to the mass eigenstates in the CP-even sector is again given by $R$
with the angles $\alpha_i$ varying in the same range as before.
The model
has 12 independent parameters: $v, \, \tan \beta, \, \alpha_{1,2,3}, \, m_{H_{1,2,3}}, m_A, \, m_{H^\pm}$ and $m_{12}^2$.
Extending the $\mathbb{Z}_2$ symmetry to the Yukawa sector we end up with the same four types of Yukawa models.
A detailed study of the N2HDM was performed in ~\cite{Muhlleitner:2016mzt}. 

\section{The 125 GeV Higgs}

\subsection{Higgs couplings to gauge bosons}

\noindent
The values of the tree-level 125 GeV Higgs couplings to gauge bosons are all smaller (in all the models discussed)
 than the corresponding SM coupling. This is a consequence of unitarity - the sum of the squared couplings $g_{h_iVV}$, where $V=W,Z$
and $H_i$ denotes one of the CP-even Higgs boson, has to be equal to the corresponding SM coupling $g_{hVV}^{SM}$. Taking the lightest Higgs in the model to be
the 125 GeV one (and just call it $h$), to make the discussion easy, the couplings to gauge bosons in the RxSM and in the 2HDM are modified relative to the SM as
\begin{equation}
g_{hVV}^{RxSM} = \cos (\alpha_1) g_{hVV}^{SM}; \qquad  g_{hVV}^{2HDM} = \sin (\beta -\alpha) g_{hVV}^{SM}  \, ,
\end{equation}
while for the CxSM, C2HDM and N2HDM the couplings are modified relative to the RxSM and to the 2HDM, respectively, as
\begin{equation}
g_{hVV}^{CxSM} = \cos (\alpha_2) g_{hVV}^{RxSM} ; \qquad  g_{hVV}^{N2HDM} = \cos (\alpha_2) g_{hVV}^{2HDM} ; \qquad g_{hVV}^{C2HDM} = \cos (\alpha_2) g_{hVV}^{2HDM}  \, .
\end{equation}
However, the angle $\alpha_2$ has very different meanings in these models, that is, it measures different contributions to the 125 GeV Higgs: the imaginary component of the singlet
in the CxSM, the singlet component in the N2HDM and the CP-odd component in the C2HDM.

\subsection{Higgs couplings to fermions}

\noindent
In the case of the singlet extensions, the 125 GeV Higgs Yukawa couplings are modified relative to SM by the the same factor
that modified the Higgs to gauge boson couplings:  $\cos (\alpha_1)$ for the RxSM
and  $\cos (\alpha_1) \cos (\alpha_2)$ for the CxSM,
\begin{equation}
Y_{hff}^{RxSM} = \cos (\alpha_1) Y_{hff}^{SM}; \qquad  Y_{hff}^{CxSM} = \cos (\alpha_2) Y_{hff}^{RxSM}  \, ,
\end{equation}
while for the N2HDM and for the C2HDM the couplings are modified relative to the 2HDM, respectively, as
\begin{equation}
Y_{hff}^{N2HDM} = \cos (\alpha_2) Y_{hff}^{2HDM} ; \qquad Re(Y_{hff}^{C2HDM}) = \cos (\alpha_2) Y_{hff}^{2HDM}  \, .
\end{equation}
That is, they are modified exactly like for the gauge bosons, except for the C2HDM for which there is an imaginary component 
of the Yukawa coupling that may have one of the following forms
\begin{equation}
Im(Y_{hff}^{C2HDM}) = \pm i \, \frac{\sin (\alpha_2)}{\tan \beta} \, Y_{hff}^{SM}; \qquad Im(Y_{hff}^{C2HDM}) = \pm i \, \sin (\alpha_2) \, \tan \beta \, Y_{hff}^{SM} \, .
\end{equation}
depending on the model type (see~\cite{Fontes:2015mea, Fontes:2017zfn} for details). As discussed in~\cite{Fontes:2015mea}, even if the angle that measures
the amount of CP-violation $\alpha_2$ is small, the pseudoscalar component can still be large if $\tan \beta$ is large.

\subsection{Bounds on the $h_{125}$ components}

\begin{table}[h!]
\begin{center}
\begin{tabular}{lcccccc} \toprule
Model & CxSM & C2HDM II & C2HDM I & N2HDM II & N2HDM I & NMSSM \\ \midrule
$\left(\Sigma\, {\rm or} \,\Psi\right)_{\text{allowed}}$ & 11\% & 10\% & 20\% & 55\% & 25\% &
 41\% \\
\bottomrule
\end{tabular}
\caption{Allowed singlet and pseudoscalar (for the C2HDM) admixtures after the LHC Run 1. \label{tab:admixtures}}
\end{center}
\end{table}

\noindent
In the models discussed in this work, the 125 GeV Higgs can be a combination from the two doublets like in 2HDM
it can have a CP-even and a CP-odd admixture as in the C2HDM or it can have a singlet admixture as in the singlet extension
and in the N2HDM. Using the ATLAS and CMS combined measurements~\cite{Aad:2015zhl} of the Higgs couplings after the LHC Run 1,
we can derive the maximum allowed admixtures~\cite{Muhlleitner:2017dkd} which are shown in 
table~\ref{tab:admixtures} \color{black}, where $\Sigma$ ($\Psi$) stands for the singlet (pseudoscalar) admixture of the 125 GeV Higgs. Results
are shown for a few selected models including the Next-to-Minimal Supersymmetric Standard Model (NMSSM). It is
clear from the table that substantial admixtures are still allowed after Run 1. However, in a future
electron-positron collider such as CLIC, the precise measurements of the couplings will reduce the admixtures
well bellow the percent level. Using the CLIC predictions for the measurements of the Higgs couplings~\cite{Sicking:2016zcl, Abramowicz:2013tzc}
we found that~\cite{Azevedo:2018llq}  the bounds on the admixtures are completely dominated by the measurement of $ \kappa_{HZZ}$
for $\sqrt{s} = 350$ GeV and a luminosity of 500 fb$^{-1}$ and by $\kappa_{HWW}$ for  $\sqrt{s} = 3$ TeV and a luminosity of + 2.0 ab$^{-1}$,
where $\kappa_{Hii}^2 = \Gamma_{Hii}^{BSM}/\Gamma_{Hii}^{SM}$. With very precise measurements of  $\kappa_{ZZ,WW}$ and because
the unitary relation~\cite{Azevedo:2018llq}
\begin{equation}
  \kappa^2_{ZZ,WW} + \Psi + \Sigma \leq 1\,.
\end{equation}
holds in all models and is independent of the Yukawa type, the bounds on the admixtures 
 (assuming that the central values are the SM predictions), will be roughly the same for all models and are given by~\cite{Azevedo:2018llq}
\begin{itemize}
  \item $\sqrt{s} = 350$ GeV and a luminosity of 500 fb$^{-1}$:   $\Sigma, \Psi < 0.85\%$ from $\kappa_{HZZ}$
  \item $\sqrt{s} = 3$ TeV and a luminosity of + 2.0 ab$^{-1}$: $ \Sigma, \Psi < 0.30\%$ from $\kappa_{HWW}$ \, .
\end{itemize}


\section{Non-SM like scenarios}

\noindent
There are many phenomenologically interesting scenarios for the extended Higgs models. New scalars are predicted and particularly charged
Higgs bosons which would definitely signal new physics beyond the SM. The most interesting signals which would change our view
of the scalar sector are given by the C2HDM. In fact, as discussed in~\cite{Fontes:2015xva, King:2015oxa}, if a new Higgs is found with substantial
decays to $H_2 \to h_{125} Z$ and  $H_2 \to Z Z$ when combined with the already observed $h_{125} \to Z Z$ it would
strongly hint a CP-violating scalar sector. However, only a detailed investigation of the model could confirm the CP-nature
of the new sector because decays of the type $A \to ZZ$ are induced at one-loop in CP-conserving scalar sectors.

\begin{figure}[h!]
  \centering
  \includegraphics[width=0.77\linewidth]{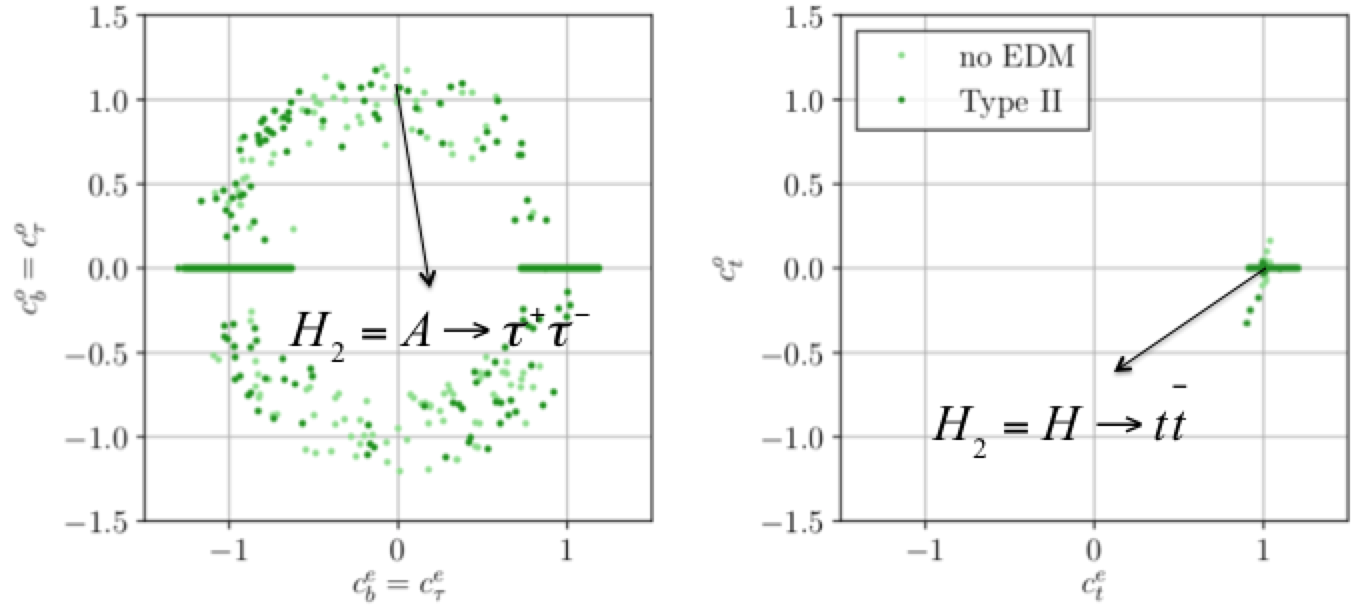}
  \caption{Allowed points in the Type II C2HDM for the case when $H_2$ is the 125 GeV Higgs. We show the points
  in the plane odd versus even Yukawa couplings. Left: $b$ and $\tau$ couplings; right: $t$ couplings. The Lagrangian
  is written as proportional to $ \bar{\psi}_f \left[ c^e(H_i ff) + i c^o(H_i ff) \gamma_5 \right] \psi_f H_i $.
    }\label{fig:1}
\end{figure}

\noindent
CP-violating sectors also allow for peculiar situations such as the one described in figure~\ref{fig:1} \color{black}.  We present
the allowed points for the Type II C2HDM for $H_2 = 125$ GeV. In the left panel points that are pure pseudoscalar
are still allowed (in $b$ and $\tau$ couplings) while in the right panel we see that only points with a very small 
pseudoscalar component are allowed. Therefore, if direct detection concludes that the Higgs is mostly a scalar in the ttH coupling
but it is mostly a pseudoscalar in the $\tau \tau H$ coupling, this can be a sign of CP-violation. 

\noindent
Finally we note that a decay of a scalar into two other scalars of different masses is sometimes one of
the best search channels~\cite{Costa:2015llh} for the models where these decays are allowed. This is not
possible in the 2HDM but it is possible for all other models presented in this work. Experimental searches
for these type of decays are therefore important for the next LHC Run.
All points comply with the most relevant theoretical bounds and the most up-to-date experimental results.

\section{Comparing models}

\begin{figure}[h!]
  \centering
  \includegraphics[width=0.77\linewidth]{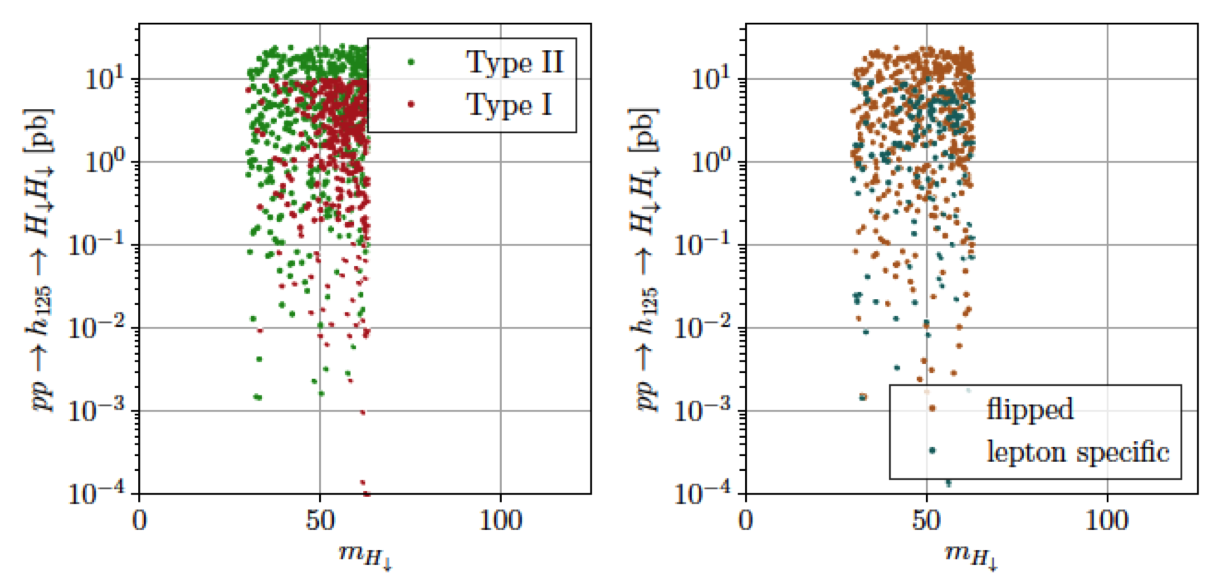}
  \caption{Total rates for the production of a $h_{125}$ Higgs decaying into two lighter scalars of the same mass.
  Left: Type I and Type II; right: Lepton Specific and Flipped.
    }\label{fig:2}
\end{figure}

\noindent
If a new scalar is found we need to understand if its properties point to a particular model or if there are models
where it is excluded. We have compared several models in recent papers
to find that event rates are sometimes enough to choose particular models in given regions
of their parameter space. Furthermore,
even the different  Yukawa versions of a specific model can sometimes be distinguished.
In figure~\ref{fig:2} \color{black} we show the total rates for the production (in $gg$+$bb$) of a $h_{125}$ Higgs decaying into two lighter scalars of the same mass.
In the left panel we show the results for Type I and Type II and in the right panel we show results for the Lepton Specific and Flipped models. We use
the notation $H_{\downarrow}$ to identify the lightest scalar (non-125) in the model. Clearly, all versions of the C2HDM can be probed at the next LHC run. Also, if the cross sections
are above 1 $pb$ some of Yukawa versions are favoured~\cite{Fontes:2017zfn}, Type II on the left and the flipped model on the right.

\section{Conclusions}

\noindent
In this work we have presented and discussed some simple extension of the scalar sector of the SM.
We have shown how the 125 Higgs can deviate from the doublet like structure by 
looking at the admixture with the singlet component and in the case of the C2HDM
the admixture with the CP-odd component. We concluded that until the end of the LHC
the bounds will be quite different in the models presented and will be of the order 
of tens of percent. 
At a future electron-positron collider such as CLIC the bounds on the admixtures become very strong (below 1\%)
and all models have roughly the same bounds due to unitarity. In such a scenario, new physics can only be seen through
the discovery of a new scalar.

\noindent
Some of the extension presented provide very interesting signals of new physics. Not only charged Higgs boson
are predicted in most models but the C2HDM is particularly interesting if certain combinations of three decays
are seen or if the CP-nature of the scalars can be studied in different channels in direct searches.

\noindent
There has been an effort to calculate electroweak radiative corrections to Higgs decays in these models
and in particular for the singlet extension~\cite{Bojarski:2015kra, Kanemura:2017wtm, Costa:2017gky}
for the 2HDM~\cite{Krause:2016oke, Denner:2016etu} and for the N2HDM~\cite{Krause:2017mal}.
Taking into account the uncertainties in those corrections and the very broad allowed parameter
space in the models, no relevant conclusions can be drawn except perhaps if a new particle
is found.

\noindent
Several codes based on {\tt HDECAY}~\cite{Djouadi:1997yw, Djouadi:2018xqq} for each
of the models presented are available for the calculation of all Higgs branching ratios,
including the state-of-the art higher order QCD corrections and possible off-shell decays:
\begin{itemize}

\item
Singlet extension, both for the RxSM and for the CxSM in their symmetric and broken phases~\cite{Costa:2015llh}, named  {\tt sHDECAY}~\footnote{The program {\tt sHDECAY} can be downloaded
  from the url: \url{http://www.itp.kit.edu/~maggie/sHDECAY}.}. 

\item
2HDM~\cite{Harlander:2013qxa} as part of the  {\tt HDECAY}  release and C2HDM~\cite{Fontes:2017zfn} named {\tt C2HDM\_HDECAY}~\footnote{The program {\tt C2HDM\_HDECAY} can be downloaded
from the url: \url{https://www.itp.kit.edu/~maggie/C2HDM}.}.

\item
N2HDM named {\tt N2HDECAY}\footnote{The program {\tt N2HDECAY} is available at: \url{https://gitlab.com/jonaswittbrodt/N2HDECAY.}
}~\cite{Muhlleitner:2016mzt, Engeln:2018mbg} which implements the N2HDM in several phases. 

\end{itemize}


\begin{wrapfigure}[7]{L}[0pt]{0.3\textwidth}
\vspace{-20pt}
  \begin{center}
    \includegraphics[width=0.22\textwidth]{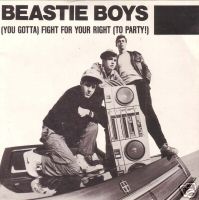}  
  \end{center}
  \vspace{-20pt}
  \caption{The right to party.}
  \vspace{-10pt}
  \label{fig:3}
\end{wrapfigure}

Finally, it can happen that the LHC will not show any signs of new physics in the next years. In that case, particle physicists can spend their
 time having fun building new models and exploring them. It could be that one of these models will finally answer the outstanding problems
in particle physics and that it will predict signatures that were not searched for so far at the LHC. It is the right of a theorist to party! (see figure~\ref{fig:3}. \footnote{Figure from wikipedia.})

%

\end{document}